\documentstyle[amssymb,epsfig]{article}
\begin{document}
\newcommand{\be}{\begin{equation}}
\newcommand{\ee}{\end{equation}}
\newcommand{\ba}{\begin{eqnarray}}
\newcommand{\ea}{\end{eqnarray}}
\newcommand{\no}{\nonumber \\}
\newcommand{\gsim}{\mathrel{\hbox{\rlap{\lower.55ex \hbox {$\sim$}}
                   \kern-.3em \raise.4ex \hbox{$>$}}}}
\newcommand{\lsim}{\mathrel{\hbox{\rlap{\lower.55ex \hbox {$\sim$}}
                   \kern-.3em \raise.4ex \hbox{$<$}}}}
\def\be{\begin{eqnarray}}
\def\ee{\end{eqnarray}}
\def\bea{\be}
\def\eea{\ee}
\newcommand{\e}{{\mbox{e}}}
\def\del{\partial}
\def\vr{{\vec r}}
\def\vk{{\vec k}}
\def\vq{{\vec q}}
\def\vp{{\vec p}}
\def\vP{{\vec P}}
\def\vt{{\vec \tau}}
\def\vs{{\vec \sigma}}
\def\vJ{{\vec J}}
\def\vB{{\vec B}}
\def\hatr{{\hat r}}
\def\hatk{{\hat k}}
\def\roughly#1{\mathrel{\raise.3ex\hbox{$#1$\kern-.75em%
\lower1ex\hbox{$\sim$}}}}
\def\lsim{\roughly<}
\def\gsim{\roughly>}
\def\fm{{\mbox{fm}}}
\def\vx{{\vec x}}
\def\({\left(}
\def\){\right)}
\def\[{\left[}
\def\]{\right]}
\def\EM{{\rm EM}}
\def\barp{{\bar p}}
\def\zz{{z \bar z}}
\def\mus{{\cal M}_s}
\def\abs#1{{\left| #1 \right|}}
\def\ve{{\vec \epsilon}}
\def\nlo#1{{\mbox{N$^{#1}$LO}}}
\def\MS{{\mbox{M1V}}}
\def\mut{{\mbox{M1S}}}
\def\Qt{{\mbox{E2S}}}
\def\rM{{\cal R}_{\rm M1}}\def\rE{{\cal R}_{\rm E2}}
\def\la{{\Big<}}
\def\ra{{\Big>}}
\def\lsim{\mathrel{\rlap{\lower3pt\hbox{\hskip1pt$\sim$}}
     \raise1pt\hbox{$<$}}} 
\def\gsim{\mathrel{\rlap{\lower3pt\hbox{\hskip1pt$\sim$}}
     \raise1pt\hbox{$>$}}} 
\def\N{${\cal N}\,\,$}

\def\J#1#2#3#4{ {#1} {\bf #2} (#4) {#3}. }
\def\PRL{Phys. Rev. Lett.}
\def\PL{Phys. Lett.}
\def\PLB{Phys. Lett. B}
\def\NP{Nucl. Phys.}
\def\NPA{Nucl. Phys. A}
\def\NPB{Nucl. Phys. B}
\def\PR{Phys. Rev.}
\def\PRC{Phys. Rev. C}

\renewcommand{\thefootnote}{\arabic{footnote}}
\setcounter{footnote}{0}

\vskip 0.4cm \hfill { }
 \hfill {\today} \vskip 1cm

\begin{center}
{\LARGE\bf  A Gravity Dual of RHIC Collisions}
\date{\today}

\vskip 1cm {\large Edward
Shuryak$^{a}$\footnote{E-mail:shuryak@tonic.physics.sunysb.edu},
Sang-Jin Sin$^{b}$\footnote{E-mail: sjsin@hanyang.ac.kr} and
Ismail
Zahed$^{a}$\footnote{E-mail:zahed@zahed.physics.sunysb.edu} }


\end{center}

\vskip 0.5cm

\begin{center}

$^a$ {\it Department of Physics and Astronomy, SUNY Stony-Brook,
NY 11794}

$^b$ {\it Department of Physics, Hanyang University, Seoul
133-791, Korea}

\end{center}

\vskip 0.5cm
\begin{abstract}

In the context of the AdS/CFT correspondence we discuss the
gravity dual of a heavy-ion-like collision in a variant of ${\cal
N}=4$ SYM. We provide a gravity dual picture of the entire process
using a model where the scattering process creates initially
a holographic shower in bulk AdS. The subsequent gravitational
fall leads to a moving black hole that is gravity dual to the
expanding and cooling heavy-ion fireball. The front of the fireball
cools at the rate of $1/\tau$, while the core cools as $1/\sqrt{\tau}$
from a cosmological-like argument. The cooling is faster than Bjorken
cooling. The fireball freezes when the dual black hole background is
replaced by a confining background through the Hawking-Page transition.

\end{abstract}

\newpage

\renewcommand{\thefootnote}{\#\arabic{footnote}}
\setcounter{footnote}{0}

\section{Introduction}

The AdS/CFT correspondence~\cite{adscft} has provided a framework
for discussing a strongly coupled regime of gauge theories in terms
of their gravity dual description. The equilibrium finite
temperature problem using a black-hole background was discussed
in~\cite{witten1} and  the bulk
thermodynamics  was discussed in \cite{thermo}.
The transport coefficients~\cite{Son} in this approach gave a result
surprisingly close to what is measured in current heavy-ion
collisions at RHIC. Also, the AdS/CFT provides a simple explanation
for high energy jet quenching at RHIC~\cite{jet} an issue of
considerable experimental interest in the
sQGP~\cite{SZ12,GML,SZADS}. IN fact, following Fermi \cite{Fermi},
who first suggested that the collision of strongly interacting
matter will produce  a thermal state, Landau \cite{landau_53}
observed that the system would follow an  adiabatic cooling path
through transiting thermal states with {\em entropy} conservation.
He further pointed out that the evolution should then be described
by (ideal) hydrodynamics. Indeed, one of the  key feature of the
'strongly interacting' Quark Gluon Plasma (sQGP) is precisely the
observation of a hydrodynamical expansion in the form of radial and
elliptic flow at RHIC.

In \cite{Giddings}, black hole formation in AdS space and its gauge
theory dual was discussed using the setting put forward by
Polchinski and Strassler~\cite{Polchinski}. More recently, in
\cite{Aharony}, the authors  discussed a scenario leading to
a long-lived or quasi-static plasma ball in the same setting.
However, high energy collisions in QCD  do
not result in stopping of the through-going partons.

The purpose of this paper is to provide a general scheme to
 address the complex issues of thermalization,
entropy formation, cooling and freeze-out in a heavy-ion collision
using the gravity dual description of the {\it time dependent}  black
hole formation. A brief summary of our results was presented
in~\cite{BROOKHAVEN}, whereby we use the AdS/CFT framework along the
lines suggested in~\cite{jet,
Polchinski,Giddings,Aharony,RSZ,JP,myers,Nastase}. In short, it should
be a process of {\em black hole formation} followed by a
Hawking-Page transition which, from the boundary point of view,
corresponds to thermalization, cooling and finally freezeout through
a confinement-deconfinement phase transition respectively. Although
the secondary scattering of partons at the boundary is a quantum
mechanical process,  its gravity dual is a classical one.

QCD is asymptotically free and at very short
distances the interaction cannot be strong. Thus Landau's scenario
can only be applicable after some `parton thermalization' time. In
this respect, strongly coupled $\cal N$=4 SUSY YM theory is
simpler since it is strong from fiat. Hence if we can
prove that Landau hydrodynamics works in this theory, perhaps with
some modifications and corrections, then we can hope to extend the
arguments to more QCD-like theories with asymptotic freedom and
chiral-deconfinement transitions.


The rest of the paper is organized as follows: In section 2,  we
give a step-by-step gravity dual to a heavy ion collision,
emphasizing the falling of  closed strings created by the scattering
before the large black hole creation.
We argue that the black hole formation is
an inevitable consequence of the falling in AdS space.
We suggest that the cooling rate of the initial stage of the fireball
relates directly to the falling rate of the black-hole in AdS.
In section 4, we describe the late stage cooling of
the fireball using the idea of brane cosmology.
Our conclusions and discussions are in section 5.
In the appendices sketches of various ideas for future developments
 are drawn.
In appendix A, we give a brief summary of some RHIC experimental
facts  for readers unfamiliar with these current experiments.
In Appendix B, we comment on the recently suggested plasma ball.
In Appendix C, we discuss collision geometries, entropy production,
and give some estimates for physical quantities such as the saturation
scale in terms of AdS black hole parameters.
In Appendix D, we discuss temperature gradients as corrections
to the homogeneous expansion described in section 4.

\section{RHIC collision and dual black hole formation}

Recently two of us have suggested \cite{jet} that real-time
dynamics such as jet quenching in RHIC has a gravity dual description
in the form of a gravitational wave falling on the black hole. The
opacity length was found to be independent of the jet energy at
strong coupling~\cite{jet}. In a related but  different picture,
Nastase suggested that 5d black holes are formed through
gravitational collisions of shock waves \cite{Nastase}, following
on the original work of t'Hooft in flat ~\cite{tHooft} and
Giddings in AdS~\cite{Giddings}. Aharony, Minwalla and
Wiseman~\cite{Aharony} suggested the black hole dual of static
plasma balls in the pure SYM context.
It would be interesting to describe the dynamical process by which
such object if any is formed.

Before proceeding further, we remark:

\begin{enumerate}
\item
A static black hole is assumed to be formed at once in the IR region
\cite{Nastase,Aharony}. In heavy ion collisions,
the fire ball takes a time (albeit short) to form.
 To describe formation
and evolution process relevant to   thermalization and cooling, the
static approach is not appropriate.

\item
The assumption of a fixed temperature for the resulting
black-hole~\cite{Nastase} is unrealistic. In heavy-ion collisions
there is no fixed temperature for the fireball, instead it undergoes
 an adiabatic path in the phase diagram.

\item The heavy-ion collision involves fundamental (quark) probes
which are embedded in the UV region~\cite{RSZ,JP} as opposed to
the adjoint (glueball) probes set in the IR region~\cite{Aharony}.

\item The bulk of the thermalization in heavy ion collision follows the
thousands of elastic collisions each transferring energy of order
$N_c^0$ as opposed to a full stopping of energy of order $N_c^2$
as in~\cite{Aharony}.

\item
The arguments presented in~\cite{Nastase,Aharony} apply
equally to proton-proton collisions for which
there is no evidence of hydrodynamics behavior,
a hallmark of a large black-hole.

\end{enumerate}

The scattering process in the boundary occur with definite
energy. How does this translate in the gravity dual space?

\subsection{Where is the holographic image of initial scattering?}

In \cite{Polchinski}, Polchinski and Strassler argued that the gauge
theory scattering amplitude is dominated by a contribution in the
dual picture stemming from the height $r_{scat}\approx \sqrt{s}$.
For definiteness, we briefly review  \cite{Polchinski}.

Consider the exclusive process $2\to m~ particles$.
For the gauge theory momentum $p^\mu$, we associate the string
theory momentum in bulk $\tilde p^\mu$ set by the height $r$
through~\footnote{Here
we change the scale of the red shift factor from the convention of
\cite{Polchinski} for later convenience. In our convention, the
gauge theory string tension $\hat \alpha'=\Lambda^{-2}$ with the
minimum height $r_{min}=R^2\Lambda$ defined by the minimum
glueball mass $\Lambda$. }

\be \sqrt{\alpha'}{\tilde
p}^\mu={R^2\over r}p^\mu.
\ee
The gauge theory  amplitude $A(p)$ at the
boundary and the string theory amplitude in a flat space $A_s(\tilde{p})$
are related by the postulated formulae
 \be
A(p)= \int dr d\Omega_{5} \sqrt{g} A_s(\tilde
p)\prod_{i=1}^{m+2} \psi_i(r,\Omega).
\ee
The string amplitude $A_s(\tilde p)$ fall off exponentially
for small $r$, and the wave functions fall off at large $r$
so that the maximum contribution occurs at finite height

\be
r_{scat}\sim R^2 p. \label{scatt}
\ee
More explicitly, for $m=2$,

\be
r_{scat}\sim R^2 \sqrt{|t| \ln (s/|t|)/(\Delta-4)} \,\,.
\ee
If $r_{scat}$ is not smaller than the IR cut-off $r_{min}$ (the position of
the IR brane) the image of the collision in bulk is located at a certain
height in AdS rather than the bottom~
as is claimed in \cite{Aharony,Nastase}. The higher
$\sqrt{s}$ the closer to the boundary~\footnote{The image, although
localized at $r_{scat}$ is not sharp unless the object has high
conformal weight, for which the size of the holographic image along
the $r$ direction, $\delta r_{scat}$,  is estimated to be \be {\delta
r_{scat}\over r_{scat}}\sim {1\over\sqrt{ \Delta }}. \ee
Therefore for high conformal weight in 4d,  the holographic image is
localized. To simplify the discussion, let's discuss the scattering
in pure SYM without quarks. If we model the initial beam as a highly
excited glueball, then the conformal weight
$\Delta=\sum_{i=1}^{m+2}\Delta_{i}$ can be considered to be large to
mimic a heavy ion-like object. This means that the holographic image
of the incoming beam has a well defined height and the
debris of the scattering should fall  under  AdS gravity. Below we
will argue that the debris form a a receding black hole within
a dynamical time of order $\Lambda^{-1}_{QCD}$ at the bottom or IR brane.
}.

\begin{figure}[t]
\centerline{\epsfig{file=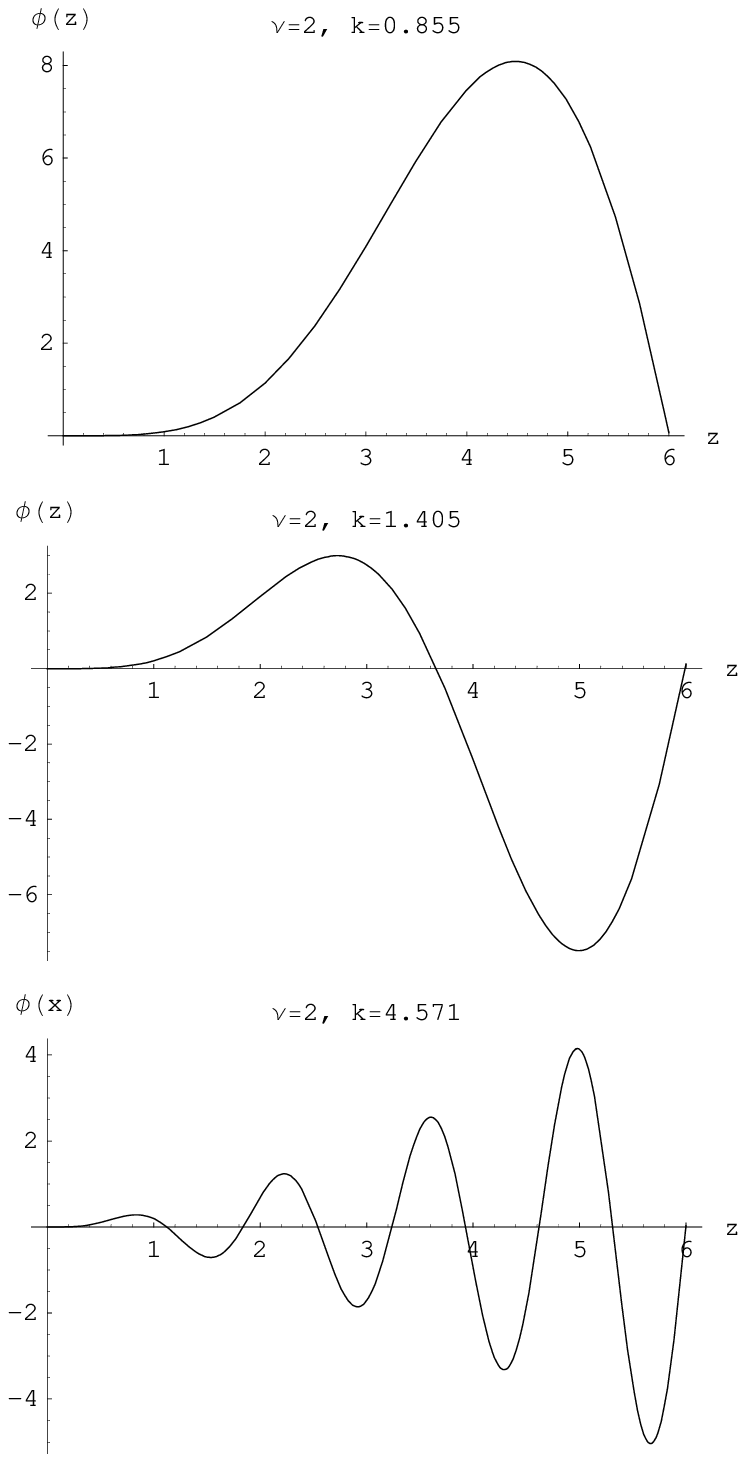,height=10cm} } \caption{\small
Wave functions of a few low energy glueball spectra. }
\label{wavefn}
\end{figure}

But how is this consistent with the fact that the AdS wavefunctions
of the incoming particles are peaked in the IR not the UV as shown
in the Fig.~\ref{wavefn}?

The answer lies in the measure factor. Since the discussions so far
did not introduce fermions (fundamental fields), we model the heavy
ion collision as a collision of two highly excited glueball states.
For an incoming glueball with definite energy, the scalar wave
function is factorized as $\Phi(x^\mu,z)=e^{ik\cdot x}\phi(z) $,
and the scalar field equation in 5 dimension,
$ (\square_5-m^2)\Phi(x^\mu,z)=0 $,
reduces to

\be
\[ z^5\partial_z
z^{-3}\partial_z -k^2z^{2} -(mR)^2 \] \phi(z)=0 ,
\ee
with $k^2=\vec{k}^2-\omega^2$. Therefore for a process with definite
energy-momentum, the wave function depends only on the mass not on
the energy. For a confining theory, we need to cut off the IR part
 by hand (restrict to $z>z_m$) or by a pertinent metric structure,
 hence we are interested in a wave function that is regular near
 the boundary($z\sim 0$), where, the wave function
behave $\phi\sim z^{2\pm\nu}$ with $\nu=\sqrt{4+(mR)^2}$ and
$R^4=4\pi g_s N\alpha'^2$. The non-normalizable part
($z^2K_\nu(kz),z^2N_\nu(kz)$) should be interpreted as the two point
function with a source at the boundary\cite{bala1}, so that its
coefficient is the strength of the source. For the initial
beam, it is on-shell and $k^2<0$ is the 4 dimensional mass. The
explicit normalizable wave function is~\cite{gkp}

\be \phi(z)= z^2
J_\nu(kz) ~~for~~ k^2 <0.
\ee
On the other hand and between the
collisions, the particles are off mass-shell ($k^2>0$)and  $k$
should be interpreted as the momentum transfer. The wave function
is

\be \phi(z)= z^2 I_\nu(kz) ~~for~~ k^2>0.
\ee

\begin{figure}[t]
\centerline{\epsfig{file=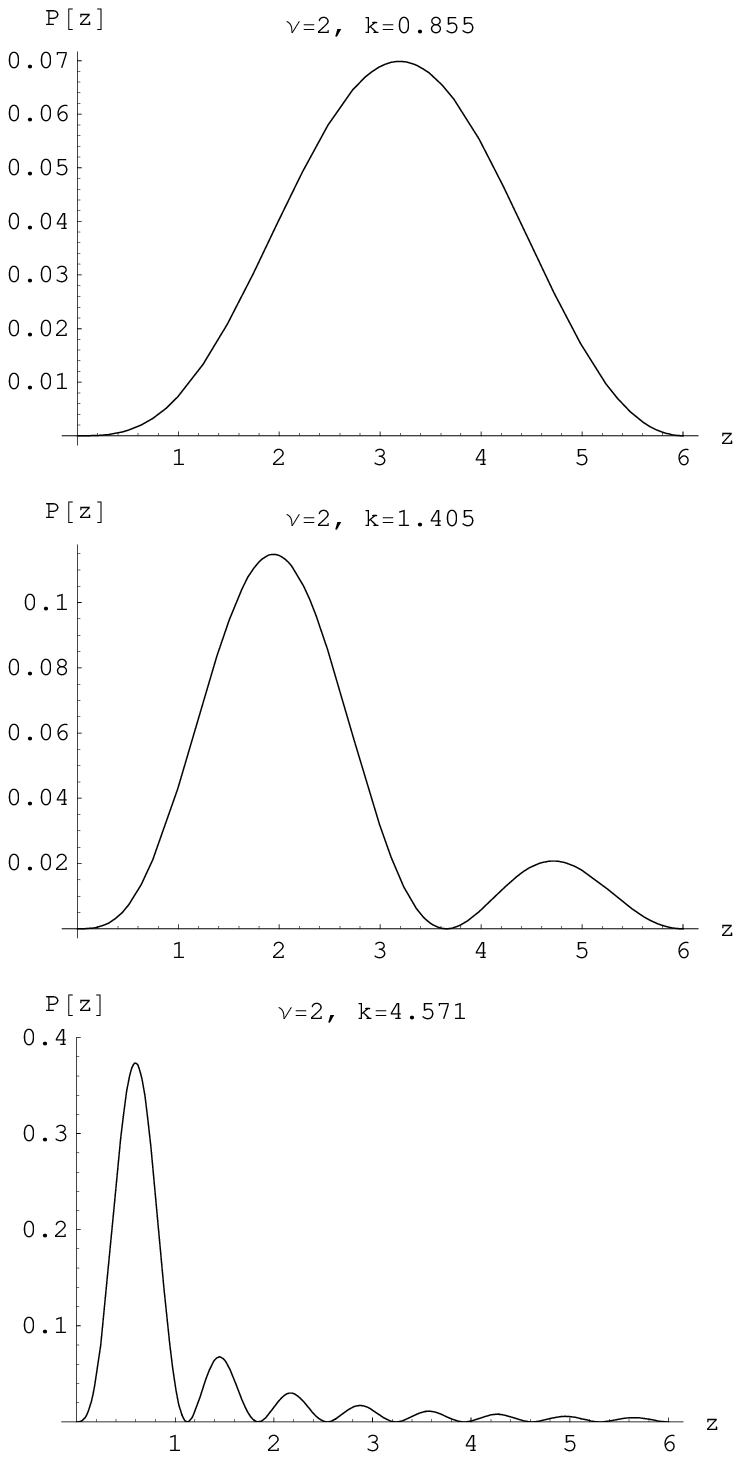,height=10cm} }
\caption{\small
Probability distributions. Notice that $z=0$ is
the boundary. For higher excitations it is more likely to be at UV region.}
\label{prob}
\end{figure}

As the 5d mass
(equivalently $\sim \nu$) increases, no qualitative change in the
wave functions is observed except that it is slightly pushed to the IR
region(larger $z$).
On the other hand, increasing the 4d mass ($k^{2}$) includes
more nodes in the allowed region and effectively pushes the
wave function into UV region. This  ``push-to-UV'' effect is
more dramatic if we consider the `radial' probability density
$P(z)=\sqrt{g}|\phi(z)|^2$. Due to the measure ($\sqrt{g}$),
the dominant peak is near the boundary rather than horizon.
We can estimate the location of the
dominant peak in terms of $x_{\nu 1}$, the first zero of the
$J_\nu(x)$. We suggest that the location of the holographic image
of the incoming glueball with mass $M_4$ $(=\sqrt{-k^2}:=k)$ is
given by

\be  r_0\sim \frac{1}{2}\frac{R^2}{x_{\nu 1}}M_4 \sim R^{2}k,
\ee
which is consistent with eq.(\ref{scatt}).

In summary, if we model a heavy ion as a glueball with large
4d mass, the holographic image of the initial beam is at the
height that is proportional to the mass. This is
consistent with Polchinski-Strassler's argument as detailed above.
The scattering takes place at a given height since their initial states
are localized there.

\subsection{Expansion and Thermalization}

Since RHIC  data shows that  the  fireball after collision is  a
thermalized, an AdS black hole should be formed. But then how?.
For this, we notice
that the AdS gravity has  anti-tidal force so that it has a focusing
property. Namely, two vertically separated particles in AdS bulk will
become closer  as they fall.
We have a gas of falling debris after collision.
  If one consider a local rest frame  of the fluid
  (lagrangian coordinate in fluid mechanics language),
  the common proper
time for all the particles can be used as a time coordinate.
This is similar to the treatment of Bjorken in 3+1 spacetime
by assuming LRF.
As we will see below, all the falling
 particles released
 from different heights arrive at the bottom  afther the same proper
 time, $\tau=\pi R/2$.
The unavoidable
consequence of this is that the motion of AdS fluid is like a
'cosmological' contraction leading to the final singularity
and  the dual of the fireball forms an AdS
black hole. Of course,  The dual of this contraction is the fireball
expansion in the boundary.

 To be more explicit,  consider a radial
 in-fall in AdS space:

\be
d\tau^2= \left( \frac{r}{R}\right) ^2 dt^2-\left(  \frac{R}{r}
\right)^2 dr^2.
\ee
For massless particle, the  motion is described by a null
geodesic with solution

\be
r=R^{2}/t\,\,,
\ee
 and the falling should start on the AdS boundary at $t=0$,
  which is consistent with the picture that
the free falling of massless particles in AdS is dual
to the free expansion in the boundary
 whose front surface is expanding with light velocity
 \cite{jet,mikhailov}. For massive particle it leads to

\be
\left( \frac{dr}{d\tau} \right)^2+\left(\frac{r}{R}\right)^2
  =\epsilon^2,
\ee
where $\epsilon=(r/R)^{2}{dt\over d\tau} $ denotes the energy per unit mass.
 The resulting motion is harmonic in proper time,

 \be
 r=R\epsilon \cos(\tau/R), \quad t=R/\epsilon \cdot \tan(\tau/R).
\ee
The period is $2\pi R$
which is {\it independent of the initial conditions.}
In case there is an IR brane,
 the initial difference in height $\delta r(\tau=0)=R\delta \epsilon$
 will be reduced to
to $\delta r (\tau)=R\delta\epsilon \cos(\tau/R)$ at the bottom.
 In terms of the boundary time $t$,

\be
 r=\frac{\epsilon R}{\sqrt{(\epsilon t/R)^{2}+1}} =
 R^{2}/t-(R/t)^{3}/2\epsilon^{2} +{\cal O}(t^{-5}), \label{earlyfall}
\ee
 so that the initial condition dependence
 ( that is the $\epsilon$ dependence) disappears rapidly as time goes on.
 We believe that this focusing effect plays an important role in
 the initial formation of the black hole geometry.
So eq. (\ref{earlyfall}) can be thought to
 describe the front surface of the fireball
which is not equilibrated.

\begin{figure}[t]
\centerline{\epsfig{file=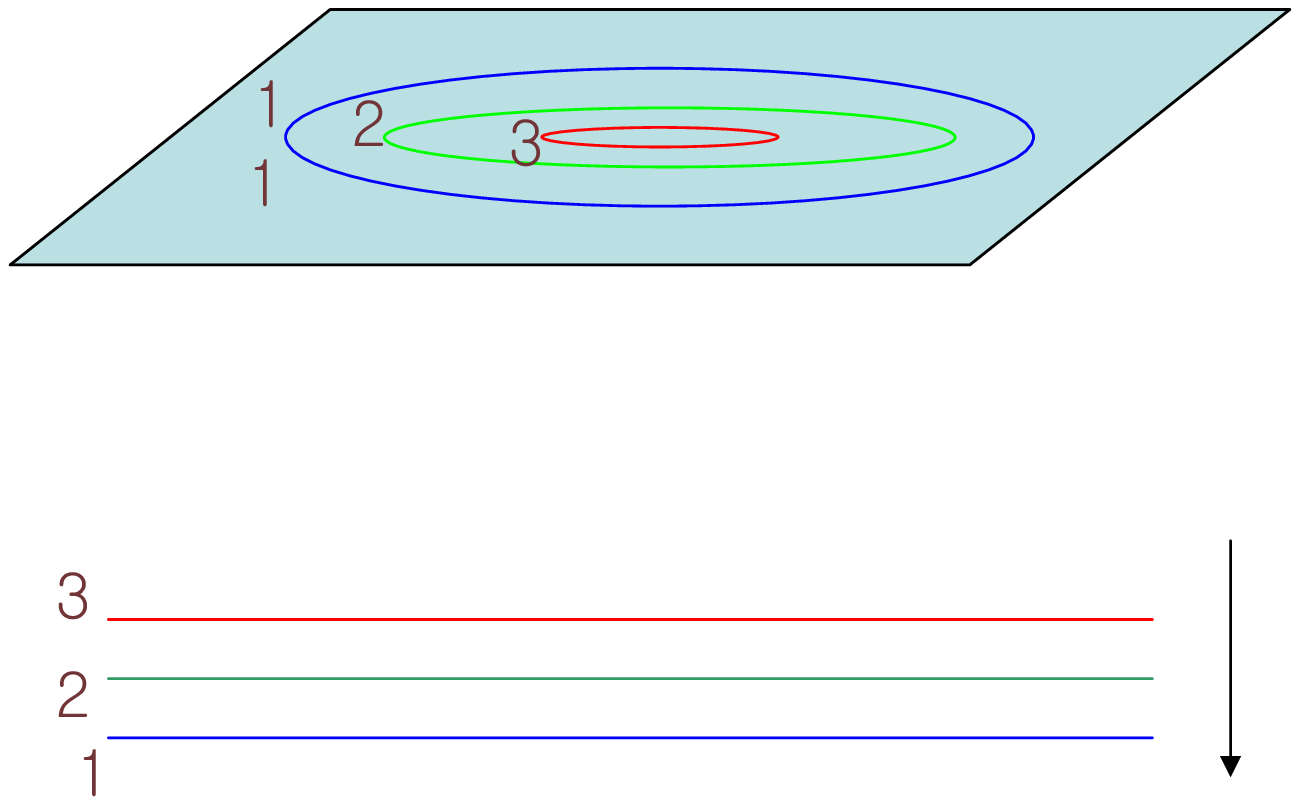,height=6cm,angle=-90} } \vskip
-1.5cm \caption{\small Holographic correspondence of the expansion in 4d
and the falling in 5d. From the boundary point of view,
the front part `1' is freely streaming
while the inner part `3' sees medium effects. From the bulk point of view:
the lower part
`1' falls freely  while the upper part `3' sees the AdS black hole
geometry. Birkhoff's theorem tells that whether the
inner part is really black hole or not is not
an issue. Thus the inner part `3' feels that it is in
thermal equilibrium.}
\label{overall}
\end{figure}

After reaching bottom (IR region) the {\it droplet} will spread
and flatten to make a pancake. For late time falling objects,
such a stack of mass on the IR brane generates a black hole
geometry due to Birkhoff's theorem. See Fig.~\ref{overall} and
its capture.
The particles inside the front surface, experience the
interaction of a medium and the expansion in the center of
the fireball is dual to the falling of a particle in the AdS black
hole background.

In the next subsection, we will consider the case with fundamental
fields (quarks).

\subsection{With quarks: creation of closed string}

Now if we have particles in the fundamental color representations
in addition to the ones in the adjoint color representation,
we need to introduce probe branes in bulk~\cite{Karch}.
A heavy meson is a
quark and antiquark connected by a string deep in AdS.
The scattering of such mesons could be realized by moving
these AdS strings, which are highly non-local objects in AdS bulk.
The collision of such strings in the bulk may form a highly distributed
object in the bulk and may not be a black hole initially.
However, the contraction and AdS fall of these objects
will give a black hole.

The mesons in N=4 theory were studied in~\cite{myers}.
They are deeply bound with mass

\be M\sim
 {m_q\over \sqrt{Ng^{2}_{YM}}}.\ee
So if we model a heavy ion by this
meson, the quark mass should be taken as large. In this case,
the holographic image of the fireball is created at a significant
height ($\sim M_q$), and falls to form a black hole at the bottom.
In this picture we can argue that for each
vertex, a closed string can be created to leave
behind a flavor brane.

\begin{figure}[t]
\centerline{\epsfig{file=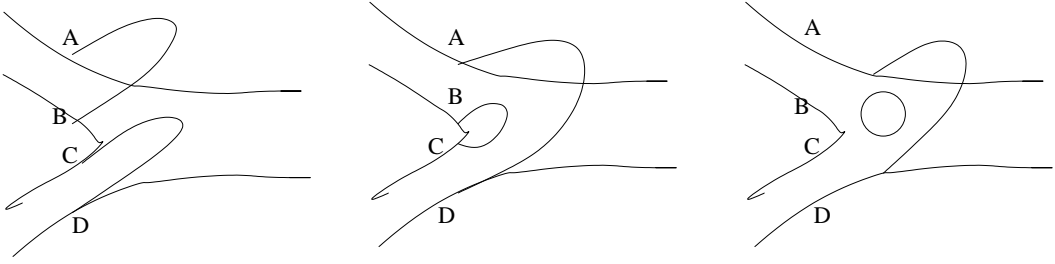,width=14cm}} \caption{\small At
each interaction vertex of two scattering mesons a closed string
must pop up. This is a unique feature of AdS space that does not
take place easily in flat space.} \label{flake}
\end{figure}

The many parton collisions at the boundary trigger {\bf i.} elastic
collisions which are dual to massive closed strings; {\bf ii.}
inelastic collisions which are dual surface flips. An example of the
former process is shown in Fig.~\ref{flake}. Since the minimal
string is not a straight line connecting two sources on the boundary
(infinite warping factor), the string must stretch inside the AdS
space~\cite{MALDA2,REY}. As two mesonic composites come together,
the recombination from $AB+CD$ to $AD+BC$ should happen just before
$B$ and $C$  touch each other, since that is energetically favored.
For example, when the separation (in boundary) of $AB$ and $BC$ are
both $L$ and that of $BC$ is $\epsilon$, then for small enough
$\epsilon$ the difference of total lengths of the string is \be
l_{AB}+l_{CD}-l_{AD}-l_{BC}=
-2\frac{c}{L}+\frac{c}{2L+\epsilon}+\frac{c}{\epsilon} > 0, \ee
where $c$ is just a constant.
Thus in a hadron-hadron scattering process, recombination of the
string must arise at the vertex (where $B$ and $C$ coincide)
generating a closed string. This is a remarkable feature of AdS
space with no analogue in flat geometry. Although the above example
is for pure AdS, we expect the mechanism to be universal regardless
of the geometry in the IR region if the UV region remains AdS.

\begin{figure}[t]
\centerline{\epsfig{file=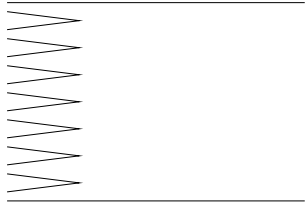,width=4cm}} \caption{\small
Multiple interaction vertices create a shower of massive closed
strings in AdS space. Some of them are mini-black holes. The strings
flake and fall towards the AdS center like a rain-fall to form
a large black hole at bottom. }
\label{flake2}
\end{figure}

Each of these liberated closed strings fall in the AdS space
under AdS gravity. Some of the the closed string states could be in
a black hole state. They merge as they fall to form a larger black
hole in the IR region as we suggested earlier.   In fact, the initial
colliding objects contain multitudes of particles, and they can not
be in thermal equilibrium at once. Therefore in the gravity dual
what forms immediately after the collision is not a single big black
hole but a multitudes of mini black holes together with other closed
string states.

The efficient creation of the particles provide a mechanism to
convert the deposited collision kinetic energy to mass resulting in
lowering the temperature scale and most of the initial energy is
deposited as mass. In real QCD, this procedure increases the
strength of the interaction by the running coupling. Here in N=4
SYM, there are no such effect since coupling does not run. From the
boundary language, the increase of number density of particle and
the increase of interaction strength is the key point to get the
efficiency in the thermalization.

\begin{figure}[t]
\centerline{\epsfig{file=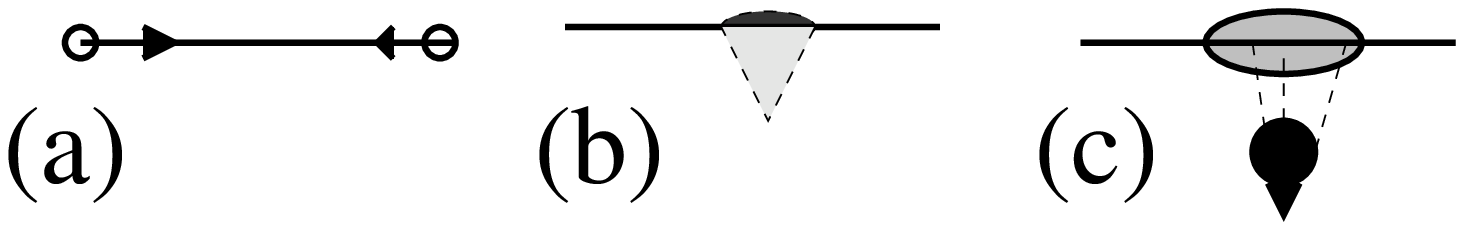,width=8cm}}
\caption{\small Schematic view of wall-wall
collision. The vertical coordinate is the 5th dimension $r$, the
horizontal one is the longitudinal one. (a)  before the collision,
  (b) shortly after the
collision: color rearrangements and
massive production of closed strings. They fall into AdS space;
(c) A larger black hole is formed at the bottom. It
will flatten to a pancake shape,  lowering its horizon.}
\label{fig_walls}
\end{figure}

The gravity dual of  the collision processes is shown schematically
in~Fig~\ref{fig_walls}. Even though matter is partially stopped on
the boundary (UV region), the initial entropy build up will cause
the formation of a fireball and its expansion. Its gravity dual is a
detached set of closed string states that fall into the AdS space.

\subsection{Final Stage: Hadronization}

As time goes on, the black-brane like object becomes thinner and
eventually unstable to density fluctuations that lead to an
instability~\cite{Gregory}. We expect that the extremely thin black
pancake will fragment into small pieces each of which evaporates
(quantum mechanically) by Hawking radiation. This  may be identified
as the confinement phase transition.
Since the resulting metric is
nothing but the AdS metric with IR cut-off, which is a confining
metric, this can be considered as the details of the Hawking-Page
transition  \cite{witten1,myers}.
\footnote{
In pure N=4 SUSY the interaction is the same
at all scales. Thus an expanding fireball of ``CFT
plasma'' will never freezeout, and will expand hydrodynamically
forever till zero temperature is reached. Freezeout can be reached
in the gravity dual by switching to confining D-brane metrics,
which is known as a Hawking-Page transition.
}
In our picture, we do not expect a quasi static black holes as in
\cite{Aharony}. This is also so in heavy-ion collisions, since not
all partons are stopped at once, the initial expansion is one
dimensional instead of three dimensional as Bjorken
suggested~\cite{BD4}.
After local equilibration is achieved in a heavy ion collision,
the matter expands and the temperature depends on both the
location and time.

With these considerations, the gravity dual of the RHIC collision
can be set up by considering the physical process together with
the general dictionary of ADS/CFT listed below.

\begin{table}[t]
\begin{tabular}{lll|}\hline\hline
gauge d=4 theory       &    string/gravity in d=10   \\  \hline
  Total cm energy per unit mass &   height of scattering brane in AdS  \\
  initial gluons, CGC  &  Aichelburg-Sexl-type shock waves  \\
  Thermalization $\rightarrow T$   & black hole formation  with $T_{bh}=T$  \\
   the entropy     & area of the horizon  \\
   rescattering ($q-q$)        & production of closed string (gravitons)   \\
   rescattering ($g-g$)        & interactions between closed strings   \\
   fireball expansion before equil. & falling b.h. in AdS bulk \\
fireball expansion after equil. &   spreading of b.h at bottom \\
   further equilibration       & merger of gravitons to black hole.  \\
   ideal hydrodynamics  & stationary black hole   \\
   hydro with  viscosity  &  growing black hole \\
    kinetic freezeout & cutoff of gravitons \\
  deconfinement & fragmentation of thin black-brane followed \\
                & by hawking evaporation \\
            \hline\hline
\end{tabular}
\caption{A vocabulary of dual phenomena in gauge and gravity
formulations.} \label{tab_cur_def}
\end{table}

\section{The cooling of the core of the RHIC fireball}

In this section we discuss cooling and
expansion of the fireball in the late stage, where the black hole
becomes a black-brane like object which is expanding in spatial direction
and lowering
its horizon continuously under the influence of AdS gravity.
The temperature decrease
is interpreted as the increase of the distance between the probe
brane and the black hole horizon.

Since the AdS space is homogeneous,
we can change the frame such that the black brane is
fixed while the probe brane is moving in the background of the
static black brane. Then the probe brane moves to the UV direction
and therefore sees
bigger scale factor of the bulk metric as it moves,
resulting in the cosmic expansion on the brane world. This
is nothing but the brane world cosmology addressed in
\cite{kraus,kiritsis,Gubser2,tye,parksin,kiritsis2}. In this way,
we identify the gravity dual of  the expansion/cooling of the
fireball as the cosmic gravitational expansion  in the
AdS-black-hole background. In other words, we approximate the
Little Bang as the Big Bang on the probe brane.
In the Little Bang, the temperature
has space and time dependence, while in the Big Bang there is no
spatial dependence, only time dependence. Therefore the approximation
is good only for the center of the fireball, which is the subject
of this section.

Although the real expansion is mostly 1 dimensional, we believe
that the thermally equilibrated expansion is 3 dimensional in nature.
This is because the 1 dimensional expansion is driven by the ultra
relativistic  motion of the initial particles whose speed can not
be caught up by the interactions.

\subsection{Big Bang on a moving brane }

We consider a class of metric given by the near horizon limit
of non-extremal $D_p$ branes:
 \be
ds^2=g_{00}dt^2+g(r)d\vec{x_p}^2+ g_{rr}(r)dr^2+g_Sd\Omega_{8-p}\,\,,
\ee
where $g =(r/R)^{(7-p)/2}$, $|g_{00}|= (r/R)^{(7-p)/2}
(1-(b/r)^{7-p})=g^{-1}_{rr}$ and
$g_S = r^2(R/r)^{(7-p)/2}$. The dilaton is given by
\be
e^{2\phi}=\(\frac{R}{r}\)^{(7-p)(3-p)/2}.
\ee

 If we neglect the brane bending effect and consider the
configuration of zero angular momentum of the brane around the
sphere, the DBI action for the $D_p$ brane \be S_p=-T_p\int
d^{p+1}\xi e^{-\phi}\sqrt{-\det{\gamma_{\alpha\beta}}}-T_p\int
C_{p+1}, \ee can be written as
\be S_p=-T_p\int d^{p+1}\xi
e^{-\phi}g^{p/2}\sqrt{|g_{00}|-g_{rr}{\dot r}^2}.
\ee
Since there
is no explicit time dependence
\be E=p\cdot q-L=
\frac{g^{p/2}e^{-\phi}}{\sqrt{|g_{00}|-g_{rr}{\dot r}^2}}-C,
\ee with $C=(r/R)^{7-p}$, is
a constant of motion.  Using the equation of motion
\be g_{rr}{\dot
r}^2+g_{00}+g^p|g_{00}|e^{-2\phi}/(C+E)^2=0,\ee the induced metric
can be written as \be
ds^2=-\frac{g_{00}^2g^pe^{-2\phi}}{(C+E)^2}dt^2+gdx^2.\ee Defining
the proper (cosmic) time $\tau$ by \be
d\tau=|g_{00}|g^{p/2}e^{-\phi}/(C+E)dt, \ee the induced metric can
be written as a zero curvature Friedman-Robertson form \be
ds^2=-d\tau^2+ a^2(\tau)dx^2,\ee where $a^2=g(r(\tau))$. The
equation of motion can be rewritten in terms of $a$ and $\tau$
\be \(\frac{{\dot
a}}{a}\)^2=\(\frac{(C+E)^2e^{2\phi}}{|g_{00}|g_{rr}g^p}-\frac{1}{g_{rr}}\)
\(\frac{g'}{2g}\)^2,\ee with $g'={dg/ dr}$.
Then, the equation of motion in terms of $a$
and $\tau$ is given by
\be \(\frac{\dot a}{a}\)^2={\(7-p\over
4\)^2}a^{2(3-p)/(7-p)}
\left[ (\frac{E}{a^4}+1)^2 -\(1-{b^{7-p}\over {R^{7-p}}}{1\over a^4 }\)
\right] ,
\ee
where we have used the
fact $C=(r/R)^{7-p}=a^4$.
Notice that the effect of the RR-flux
field $C$ is to provide a strong enough repulsion force to
cancel the confining $AdS$  gravity.

As $a\to \infty$ (late evolution), we have \be a(\tau)\approx
\tau^{(7-p)/(11-p)} . \ee The scale factor evolution $a(\tau)$
captures the cooling of the fire-ball at the boundary through its
holographic dual: \be \label{warping}
T(a)=\frac{T_{bh}}{\sqrt{|g_{00}|}} \approx
\frac{T_{bh}}{a(\tau)}, \ee with \be T_{bh}={(7-p) \over 4\pi b}
\cdot \({b \over R}\)^{(7-p)/2} \ee as the black hole temperature.
The local temperature is the  black hole  temperature observed by
the observer in the probe brane. This  is the actual temperature
of the fireball. As the brane moves away from the black hole, the
brane world (the fireball) expands and cools according to
$T(a)=T_{bh}/a$.

There are two interesting cases: $p=3$ and $p=4$. For $p=3$,
 \be
a(\tau)\sim \sqrt{\tau},\,\,\,\, T\sim {1\over\sqrt{\tau}}
\label{BD4} . \ee The reason for considering $p=4$ is that one of
its direction (say $x_4$) in a confining theory is compactified.
After the compactification the $p=4$ and $p=3$ are identical.
Without compactification $a\approx \tau^{3/7}$ which is a stronger
warping.

This result is to be compared with cooling in D-space.
Indeed, the entropy for a (perfect) gas is just $S\approx T^D\,V_D$.
For a relativistic d-space hydrodynamical expansion we expect
$V_D\approx V_{D-d}\,\tau^d$. For fixed entropy, the temperature falls
like $T\approx 1/\tau^{d/D}$. Bjorken 1-space expansion~(\ref{BD4}) corresponds
to $D=3$ and $d=1$, therefore $T\approx 1/\tau^{1/3}$. Fully 3-space
expansion corresponds to $T\approx 1/\tau$. The AdS case with
$T\approx 1/\sqrt{\tau}$ is faster than Bjorken in 1space
but slower than perfect hydrodynamical expansion in 3space.
It is like fractal with $d=D/2=3/2$.
One may summarize these result by saying
that {\it strong interactions slow down the expansion of the
fireball just as gravity does in the dual picture.}
We note that since the viscosity is quantum with
$\eta/(S/V_3)\approx \hbar/4\pi$ its effects are
not present in our estimates. Their consideration follow from
perturbation theory and are easily seen to delay the cooling.

\subsection{confinement phase transition}

When $T$ cools enough such that $T<\Lambda_{QCD}$, there must be
a Hawking-Page transition~\cite{HawkingPage} and the background
metric is replaced by
\be ds^2=(r/R)^{3/2}(-dt^2+ d\vec{x}^2 +f_2(r)d x_4^2)
+(R/r)^{3/2}(dr^2/f_1+r^2d\Omega^2_4), \ee where
$f_2=1-(r_{KK}/r)^3$ refers to the compactified direction.
Witten~\cite{witten1,myers} suggested that the transition to this
metric maybe interpreted as the confinement/deconfinement phase
transition. The equation of motion in the new background can be
calculated. Though minor, there are a few differences in detail of
the calculation, but rather surprisingly, the final outcome is
precisely the same with the substitution $b\to r_{KK}$. For $p=4$,
 \be \(\frac{\dot
a}{a}\)^2={9\over 16}a^{-2/3}\left[ \(\frac{E}{a^4}+1\)^2
-\(1-\frac{r_{KK}^3}{R^3}{1\over a^4 }\) \right] .
\ee
As we discussed before, The front factor $9/(16a^{2/3})$
disappears if $x_4$ is compactified (which is effectively $p=3$).
The phase transition point in terms of the brane position occurs
when the warping becomes $a_{F}$ at
\be T(a_{F})\approx T_{KK}\,\,, \ee where the Kaluza-Klein
temperature is given by $T_{KK}=3 r_{KK}^{1/2}/(4\pi R^{3/2})$.
Thus \be a_{F}= \frac {T_{bh}}{T_F}= \sqrt{\frac {b}{r_{KK}}} \ee
One may interpret the  phase transition as hiding of the black
hole horizon behind the KK singularity $r=r_{KK}\approx
1/\Lambda_{QCD}$. After this phase transition, hadron creation
begins which should be a dual to the Hawking evaporation of
black-brane after Gregory-Lafflame transition as we discussed at
section 3.
The fireball ultimatly freezes out when the pions
decouple.

So far the expansion is homogeneous.
This is consequence
of the assumption that the background is black brane.
In reality, temperature  has some gradient.
To discuss the spatial dependence as well as the time  of the temperature
we need the brane motion in the   background produced by an object
with finite extent like a black hole.
We give a touch to this important but difficult subject in appendix D.

\section{Discussion}

We close by summarizing. Recent heavy ion collisions at RHIC have
suggested that the released partonic matter produced is strongly
interacting in the form of an sQGP. Two of us have argued recently
that a good starting point for adressing key issues of the sQGP is
${\cal N}$=4 SYM at strong coupling. In this paper we have
suggested that the entirety of the RHIC collision process from the
prompt entropy release to the freezeout stage can be mapped by
duality to black hole formation and evolution in AdS space. In
other words RHIC little bang and cooling is dual to a cosmological
big bang with a flying black hole as a proceed.

We have provided a simple physical picture of black hole formation
and thermalization from string theory point of view.
We have suggested that due to the strong
interaction of the fireball liquid, the expansion is slower than
expected from the ideal gas model.  The cooling of the fireball is
$1/\sqrt{\tau}$ which is slower than Bjorken 3d cooling $1/\tau$.
The strong nature of the interaction slows down the expansion rate
hence the cooling is slower than expected from the Bjorken
solution. Cooling freezes when the background is replaced by
the confining background through the Hawking-page transition.

There is a clear distinction between coherent
parton-parton scattering and incoherent macroscopic (heavy-ion)
collision of large number of partons. In the former, the
scattering happen in IR region and information is conserved, which
is a hallmark of quantum mechanics, while in the latter the
scattering happens at the UV region and the information is
lost\footnote{A hypothetical {\it full experiment} with measuring
phases of all thousands of secondaries can still in principle
recover it.} and entropy is generated, perhaps to its maximal
value, typical for local thermal equilibrium. While entropy
generation maybe traced back to the incoherence due to the many
binary scattering in a RHIC heavy-ion collision, it is readily
understood in the gravity dual description: Since the particles
evaporate from the crowds, the (contracting) core is losing
information and  become a black hole. Although the lost
information will be back to the black hole (due to the AdS
gravity) and the black hole grows, the entire information is
hidden inside the horizon. This we believe is one of the simplest
explanation for entropy production at RHIC.

Non-cosmological like expansions with realistic fire ball
geometries on the boundary are more involved to analyze. We have
suggested that their asymptotic stages can be mapped on black hole
perturbation theory resulting in non-ideal hydrodynamics from
conventional Einstein gravity. We will report on these issues and
others in future.

\vskip 1cm

\noindent{ \large \bf Note added:} Since submitting our paper, a
new paper by Janik and Peschanski \cite{Janik2} appeared. The
cooling of the fireball is discussed using the asymptotic solution
for the metric induced by the energy momentum stress tensor set on
the {\it probe brane}. In the gravity dual, the black hole moving
horizon reproduces Bjorken's scaling for a one dimensional
expansion, in support to our arguments.

\vskip 1cm

\newpage

{\bf\noindent \Huge Appendices}

\appendix

\section{Elements of RHIC physics }

This discussion is intended to be elementary to shorten up the
vocabulary gaps between the string community and the heavy ion
community interested in the gauge-gravity problems through the
AdS/CFT correspondence.

{\bf Collision:\,\,} Experimentally we use the heaviest (and fully
ionized) nuclei (mostly $Au^{197}$ at RHIC) with as large energy per
nucleon as possible (the relativistic gamma factors $\gamma\sim 100$
in center of mass, to be increase further at LHC soon.)

One may ignore the complexities of nuclear physics and QCD
evolution, and focus solely on the partonic wave function of hadrons
or  nuclei before the collision. More precisely, as coherence is
lost anyway, one needs to know the mean $squared$ amplitudes of the
pertinent harmonics of the comoving gluon field with the so called
saturation scale $Q_s$ or equivalently the transverse density of
partons $Q_s^2$. At RHIC $Q_s$ is about 1.5 GeV for a typical
Feynman $x=10^{-2}$. It will be higher at LHC say $Q_s=6-8$ GeV at
lower $x$. A model currently used to describe the low-x part of the
nuclear wavefunction prior to the RHIC collision is the color glass
condensate (CGC). It is rooted on a weak coupling argument in QCD
contrary to what is stated in~\cite{Nastase}.

{\bf Equilibration:\,\,} This is a transition from the CGC to
thermal quarks and gluons. Solutions of classical Yang-Mills, both
for random fields~\cite{RAJU} and sphalerons \cite{SPHA} have
actually produced thermal-looking spectra but more is to be
understood, perhaps along the discussion of plasma
instabilities~\cite{YAFFE}.

{\bf Hydrodynamics:\,\,} This is a key aspect of RHIC physics.
Maintaining collective flow for systems containig just $\sim
100-1000$ particles is a nontrivial issue~\cite{hydro}, and would
not happen for usual liquids like water. Thus the  matter produced
at RHIC is now refered to as a strongly coupled quark-gluon plasma
(sQGP) or liquid. Indeed it exhibits both bulk thermodynamical
parameters and transport coefficients  (viscosity) that are
surprisingly close to what the AdS/CFT correpondence predicted for
strongly coupled N=4 SUSY YM theory. The short time behavior of the
hydrodynamical expansion is close to the 1-d Bjorken regime whereby
the temperature depends only on the proper time
$\tau=\sqrt{t^2-z^2}$. For central collisions the expansion becomes
axially symmetric before turning to a full 3d spherical expansion.
For non-central collisions there is azimuthal anisotropy which is
successfully described by hydrodynamics.

{\bf Freezeouts:\,\,} This corresponds to chemical and thermal
freezeouts whereby the change in the composition is turned off
(chemical) and the particles decouple (thermal) with free streaming.
Both freezeouts follow from the same condition $\nu_{expansion}=
\nu_{reaction}$, where we have used the covariant definition of the
expansion rate $\nu_{expansion}=\partial_\mu u^\mu$.

In cosmology, the expansion is so slow that not only strong ($pp$)
scattering survives, but even weak equilibrium through
$p+e\leftrightarrow\nu+n$ does, untill $T\approx 1$ MeV. Photons
freezeout at much lower temperatures $T\approx 0.1$ eV. At RHIC
chemical freezeout corresponds to the end of particle changing
reactions such as $2\pi\rightarrow 4\pi$, while kinetic freezeout
corresponds to the last elastic collision such as $2\pi\rightarrow
2\pi$. Experimentally both freezeouts are reasonably well measured,
the former from matter composition while the latter from particle
spectra~\cite{STARWHITEPAPER}. While the critical temperature in QCD
$T_{ch}\approx 176$ MeV is independent on the collision centrality,
the freezeout temperatures depend on the system size. For instance,
the kinetic freezeout temperature $T_{\rm kin}$ does depend on the
system size, and goes down for the largest fireballs (central
collisions) to about 90 MeV. Thus the whole range of temperatures at
RHIC is about 4-fold, from the initial $T_i\approx 350$ MeV to the
kinetic freezeout $T_{\rm kin}\approx90$ MeV. The energy density
changes by about 2 orders of magnitude.

The main reason for the rapid freezeout of a hadronic gas is the
Goldstone nature of the pions. The self-interaction through
derivatives makes it difficult to generate soft pions. At low
temperature, the pion gas collision rates can be calculated from the
leading chiral interaction (Weinberg-Tomozawa). Specifically, the
elastic rate is ~\cite{Shu_pions}

\be \nu_{\pi\pi}={T^5 \over 12 f_\pi^4} \ee The strong T dependence
follows from dimensional arguments. The inelastic rates can be found
in~\cite{GGL_89}.

At RHIC detailed numerical calculations show that the proper time
spent in the sQGP phase ($T> T_c$) the ``mixed phase'' ($T\approx
T_c$) and the hadronic phase ($T< T_c$) are all comparable. However
at LHC the sQGP should dominate. For simplicity, we may ignore the
complications inherent to the running coupling in QCD, the
confinement-deconfinement transition and the pion dynamics by
restricting the discussion to the early phase of the collision
dominated by the sQGP. If the latter phase is close to strongly
coupled N=4 SUSY matter at finite temperature, as two of us
discussed recently~\cite{jet}, it is then useful to use the duality
insights to bear on the bulk and kinetic properties of the sQGP.

\section{ Comments on plasma bubble}

The formation of a metastable ``plasma bubble" was recently discussed
  by Giddings \cite{Giddings} and its boundary and slow evaporation of
  glueballs was discussed in \cite{Aharony}.
   In QCD the idea that near the phase transition there can be a
near-stable fireball was discussed since 1970's, see in particular
\cite{ShuZhi}. Although at finite $N$ the pressure is always finite,
the ratio to the energy density $p/\epsilon$ has a minimum called
``the softest point'' \cite{HS}. If the system is produced at such
conditions the produced fireball would be especially long-lived: see
detailed hydrodynamical studies in \cite{HS}. Experimentally there
are indirect hints that fireball lifetime is indeed maximal around
collision energy $\sqrt{s}\approx 6\, GeV*A$, which unfortunately
was not studied in detail yet: there are proposals to run RHIC at
such a low energy to verify that.
  The gravity dual of such slow evaporation is
   Hawking radiation in the (usual not extra) spatial direction.
  This phenomenon is known in QCD as a ``long-lived fireball".
It was first related with the MIT-type bag model, in which a meta-stable
zero pressure state is possible. It was suggested as early as 1979 as a
possible explanation for why the secondaries produced in pp collisions
at ISR have thermal looking spectra for transverse momenta
without hydro-expansion \cite{SZhirov}.
Hydrodynamical transverse expansion
was observed  in heavy ion collisions in 1990's
in AGS experiments in Brookhaven:
its small magnitude was  attributed to the so called ``softest
point" of the Equation of State, the minimum of p(e)/e  right
after the phase transition, which is close to the initial conditions
in those collisions\footnote{Now there is a proposal under
evaluation to run RHIC at very low collision energy to possibly
produce such a fireball close to its critical state and study the
``softest" regime of a strongly coupled quark-gluon plasma experimentally.}.

Aharony et al~\cite{Aharony} argued that since the critical pressure
$p_c\approx N_c^0$ while the energy density on the plasma side of the
transition is $e\approx N_c^2$, their ratio is $1/N_c^2$,
vanishing in the limit of a large number of colors. Furthermore,
they argued that a small but nonzero critical pressure can be
further compensated by a surface tension $\sigma$, making stable
plasma drops provided that the drops are small enough\footnote{This
may explain why small systems, such as those produced in pp, do
not show hydro-expansion while large ones produced in
central PbPb collisions at similar energies do.}, with a size
$R<Rc\approx \sigma/p_c$. For large $N_c$ this can again be large since
$\sigma\approx N_c^2$.

One important feature of the arguments presented in~\cite{Aharony}
is that the evaporation of glueballs from the surface of such plasma
droplets leads to a reduction of the radius $R$ in the ratio $\sigma/p_c$,
so that inside the droplet the temperature should actually $increase$.
This is indeed qualitatively similar to what happens with an
evaporating black hole, emitting  Hawking radiation.

While interesting, this however is a doubly tuned situation, which
is rather different from what we expect at RHIC. In this case,
there is a violent expansion and cooling of the fireball as we have
attempted to describe in this paper. In our setting QCD matter
(and its collisions) are placed as a test brane removed from the
IR limit of AdS. The expansion and cooling are both due to
a warping of the metric induced by a departing black hole
albeit far towards the IR.

\section{A few estimates}
\subsection{Collision geometries}
\label{sec_geometries}
To simplify  the initial geometry of the problem, imagine that
colliding bodies may have infinite extensions in some directions,
with the solution naturally independent on the corresponding
variables. Let us call the number of ``non-contracted'' variables
$\tilde d$.

The simplest geometry (i)  would be a spherical collapse:
 One may imagine a spherical shell of matter
 collapsing into itself with
an initial radial velocity $v$ and Lorentz factor
$\gamma=1/\sqrt{1-v^2}$.
 A fireball which   is produced in this case is expanding in a
 spherically
symmetric way, producing
 a ``Little Bang'' like at RHIC, only in a
much simpler spherical geometry\footnote{This has been considered by
one of us many years ago \cite{EDMANY} for e+e- collisions and prior
to QCD. However, due to asymptotic freedom this condition cannot be
created experimentally: e+e- collisions in fact result in 2 jets,
propagating from the collision points in random directions rather
than a spherical expansion.}
 The next geometry (ii) to consider  is a collapse of a cylindrical shell,
leaving one ``non-contracted'' variable, $\tilde d=1$. The gravity
dual to it should have a black hole with one less dimension. The
geometry (iii) with  $\tilde d=2$, is a collision of two infinite 2d
walls. This is close to what happens at RHIC, where the colliding Au
nuclei are Lorentz contracted by a factor hundred into two thin
pancakes. A variant of this are light-like wall-wall collisions that
pass through each other causing surface/string rearrangements in the
minimal impact parameter region much like the parton-parton
scattering approach originally suggested in~\cite{RSZ,JP}.

The realistic~\footnote{The gauge theory under consideration is
still not QCD but a strongly coupled $\cal N$=4 SUSY YM} case (iv)
corresponding to RHIC is a collision of finite-size objects
(although as large as practically possible). Due to relativistic
boosts the nuclei get flatten in the collision direction $x_3$.
Furthermore,
for non-central collisions the overlap region is not  axially
symmetric but has an almond-like shape. Its gravity dual presumably
would create a black hole with a horizon of some ellipsoidal shape,
with different dimensions in all directions.

\subsection{Fermi-Landau model and the entropy formation}

In QCD and other asymptotically free theories we know that at small
distances (close to the origin) the interaction is weak. In the
collision the constituent partons  would literally fly through each
other. Thus the issue of entropy formation at RHIC is complex and,
as one may have suspected, not unanimous.

In contrast, in strongly coupled N=4 SUSY YM theory, there is no
relation between the coupling and the scale. At strong coupling, one
may think  that  the colliding matter is stopped, and that most of
the entropy is produced promptly at this stage. Thus, we use for
this case the Fermi-Landau (FL) model \cite{Fermi,landau_53} as a
benchmark for further comparison.

The main assumption of FL is that matter can be stopped in a
Lorentz-contracted size $R=R_0/\gamma$, where $R_0, \gamma$ are the
original size of the colliding objects and their Lorentz factor. The
volume in which it is supposed to happen is \be V\sim R_0^3
/\gamma^{3-\tilde d}  \ee where $\tilde d$ is the number of
``non-contracted'' coordinates introduced in the preceeding
subsection. The first step of the argument is to evaluate the
temperature at this  stopped stage. The energy density is \be
\epsilon = E/V\sim \gamma^{4-\tilde d}\sim T^4 \ee where the last
equality is from the EoS of matter. Therefore the temperature grows
with the collision energy as \be T\sim \gamma^{1-\tilde d/4}\ee The
next step gives the amount of entropy produced: \be S\approx T^3
V\approx  \gamma^{\tilde d/4} \label{SLANDAU} \ee One can see, that
in the  spherical collapse (i), there is no entropy growth because
$\tilde d=0$. The lesson from it is that only the cases with less
trivial geometry provide some interesting predictions.

Despite the differences between the FL model and QCD, the entropy
prediction for the  wall-on-wall case (iii), $S\sim \gamma^{1/2}\sim
s^{1/4}$, agrees with the observed multiplicity growth quite well.
We will return to the discussion of this point later.

\subsection{black hole formation}

For simplicity, we assume that the black-hole has been formed but
that the flaking of closed strings is still taking place, and ask:
under what conditions the flaking strings can be captured by the
black-hole? what is the typical accumulated entropy? what is the
typical time for this entropy formation? Most of the arguments in
this subsection are heuristic.

>From the AdS black hole metric we have
 \be ds^2=-fdt^2 +f^{-1}dr^2
+r^2d\Omega_5^2,\,\,  ~{\rm with}~\,\, f=1-G_5M/r^2+(r/R)^2. \ee
where we have ignored the distorsions caused by the boundary brane
on the black-hole. The horizon size of the black hole is \be
r_{bh}=R\left(\left(\frac{G_5M}{R^2}+{1\over4}\right)^{1/2}-
{1\over2}\right)^{1/2}. \ee Hence $r_{bh}=(R^2G_5M)^{1/4}:=b$ for a
large black hole, while $r_{bh}=\sqrt{G_5M}$ for a small black hole.
\footnote{ Large black hole means  $G_5M/R^2\gg 1$ with $f\sim
(r/R)^2(1-b^4/r^4)$, and small black hole means $G_5M/R^2\ll 1$ with
$f\sim 1-G_5M/r^2$.} The temperature of the large black hole is
given by $T_{bh}={b}/{\pi R^2}$, while that of the small black hole
is $T_{bh}=R/2\pi b^2 \sim 1/\sqrt{G_5M}$.
 The large AdS black hole does not evaporate while small
black holes can. However, the hawking temperature goes up as it
evaporate while the fireball cools as it evolve. Therefore small
black hole  seems to be improper to describe the RHIC fireball.
Therefore throughout this paper we identify the RHIC fireball with
large AdS black hole.

We can express the mass and entropy in terms of Hawking's
temperature \be M=R^6 T^4 /G_5 ,  \;\;
\frac{S}{V_3}=\frac{\pi^3}{2}\frac{R^2\,b^3}{G_{10}} \approx T^3 .
\label{critical} \ee On the other hand, using $
G_{5}^{-1}=G_{10}^{-1}R^5=M^8_p R^5,\;\; M_p={1/ l_s g_s^{1/4}},\;\;
R^4=g_s N_c l_s^4, $ we can express Hawking's temperature in term of
mass \be T=\frac {b}{\pi\,R^2} \approx \frac 1{\pi\sqrt{N_c}}\,
\left(\frac{M}{R^3}\right)^{1/4}. \ee Since $M\approx N_c^2$ (see
below) then $T\approx N_c^0$. The time the entropy is reached
corresponds to typically the falling time in AdS space \be \tau=
{\pi\over 2}R\ee Which is of order $N_c^0$. This is about the time
it takes the final black hole to reach the {\it bottom} of AdS.

In a typical RHIC experiment, hundreds of nucleons or thousands of
quarks are involved as shown above. As the interaction is almost
simultaneous, thousands of interaction vertices are involved and a
shower of massive closed strings are created and fall into the
center of the AdS space. Let $N=\pi\,R_N^2\,Q_s^2$ be the number of
such collisions with $Q_s^2$ their transverse density and
$\pi\,R_N^2$ the transverse nucleon size. In field theory charged
quantas moving with rapidity $Y$ are surrounded by extra quanta
distributed at smaller rapidities $dy=dx/x$. In QED the
Weiszacker-Williams approximation yields a flat distribution of
these quantas versus $y$, i.e. $dN/dy$ constant. In QCD $dN/dy$ is
not constant and behaves approximately as $e^{\alpha(t)(Y-y)}$. HERA
data suggests $\alpha\approx 1/4$ at $t\approx-1\,{\rm GeV^2}$. In
weak-coupling the BFKL approximation gives $\alpha_{BFKL} (0)
=(4\alpha\,N/\pi)\,{\rm ln}\,2$, while at strong coupling arguments
based on AdS/CFT duality yield~\cite{JP} $\alpha_{AdS}(t)\approx
7/96+ 0.23\,t$. In our case, we will use a transverse parton density
$dN/dydx_\perp=Q_s^2(y)$ with $Q_s$ in general y-dependent.

Since the 5th AdS coordinate $r$  is orthogonal to the boundary
collision axis, only momenta of closed strings with typically  $Q_s$
are relevant. Then the typical total energy of the closed strings is
$M\approx \int dy dN/dy\,Q_s$, with all strings assumed to be
created instantaneously at the impact. The energy of the closed
string must be identified as the radial coordinate  in the AdS
space. The strings flake towards the center of the curved AdS space
under gravity and arrive at the central region simultaneously. The
average energy per string is $\epsilon\approx Q_s$. The total energy
is therefore of order $NQ_s$. We note that $N\approx N_c^2$ and
$Q_s\approx N_c^0$, so that the total energy is of order $N_c^2$. A
black-hole forms when the horizon radius is bigger than the size of
the closed string. For the large black hole, the horizon distance is
$r_{bh}= (R^2G_5 M)^{1/4}$, where $G_5$ is the 5 dimensional
Newton's constant. Hence the black hole formation condition is
\be Q_s^{-1}\leq (R^2G_5 M)^{1/4}\,. \label{cond} \ee In terms of
$N(s)$ the number of pair collisions at the boundary, (\ref{cond})
reads
\be N \geq R^{-2}G_5 Q_s^{-5}. \ee We recall that $M\approx Q_s^3$
and $N\approx Q_s^2\approx s^\alpha$. The entropy generated at the
surface by the RHIC collision depends on the collision energy as
follows
\be S\approx T^3\,V_3\approx s^{9\alpha/8-1/2} \label{SADS} \ee or
an initial temperature $T\approx s^{9\alpha/8}$.

\section{A correction to the big bang picture}

\begin{figure}[t]
\centerline{\epsfig{file=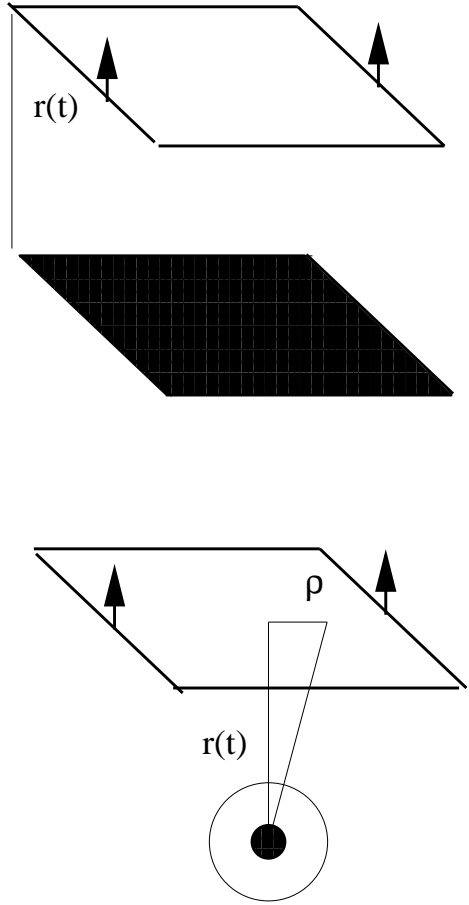,width=12cm,angle=-90}}
\caption{\small Sketches of the brane motion for the Big Bang (a)
and Little bang (b) geometries. (a) refers to a brane moving away
from large ``black brane'' with a time-dependent distance in 5-th
dimmension $r(t)$.
 (b) refers to an
asymptotically flat brane with  the brane to black hole distance
$r(t)$, while $\rho(t)$ is the effective spherical size of the
fireball on the brane. The Howking radiation has homogeneous
time-dependent temperature in (a), while it depends on $\rho$ in
case (b). \label{fig_final} }
\end{figure}

The  situation is schematically illustrated in Fig.~\ref{fig_final}
(b).

\begin{figure}[t]
\centerline{\epsfig{file=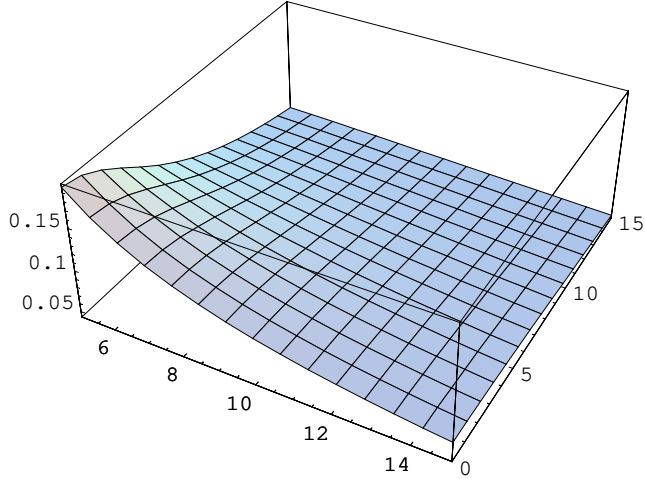,width=10cm}}
\caption{\small Plot ofthe  temperature for $0\le \rho \le
15$ as the brane position  $r_\perp(t)$ moves from 5 to 15.
The bending of the brane is ignored assuming that the brane is far from
the black hole. We have used $G_5M=1$, $R=1$, $T_0
\sqrt{f(r_0)}=1$. Notice that the warping  factor ($\sim r^2$),
which will overcompensate the apparent shrinking in the fireball size,
is not taken into account in the plot. \label{profile} }
\end{figure}

   The temperature at the center of the fire
ball is given by the warping factor as determined above using the
distance $r$ from the black hole center to the fire ball center.
The temperature at all other points in the moving brane is warped
further since the effective distance now is
$\sqrt{r_\perp(t)^2+\rho^2}$, and vanishes asymptotically.
While the precise determination of the imbedding of the moving brane is
very involved, its shape in the region far from the black hole can
be readily discussed approximately by neglecting the bending
effect.  More precisely, from the metric \be
ds^2=-fdt^2 +f^{-1}dr^2 +r^2d\Omega_5^2,\,\,  ~{\rm with}~\,\,
f=1-G_5M/r^2+(r/R)^2, \ee the temperature is approximately given by
\be T(t)=T_0 \sqrt{f(r_0)}/\sqrt{1-G_5M/(r_\perp(t)^2+\rho^2)+
(r_\perp(t)^2+\rho^2) / R^2}, \ee where $T_0$ is the
black hole temperature at a reference distance say $r=r_0$. Now,
the temperature depends both on the spatial size
$\rho$ of the fire ball on the boundary, and the distance to the black hole
$r_\perp(t)$. This is the temperature profile which has a
peak at the fireball center, $\rho=0$, and decreases  as $\rho$
increases. See figure \ref{profile}.
  The upshot of this analysis is that the presence of a distance
black hole in bulk produces small additional forces on moving
matter in the moving brane. Even without considering the bending
effect of the brane, the presence of blackhole is detected through
the metric. For instance, hydrodynamical flow of matter on the
brane is now described by $ T^\nu_{\mu;\nu}=0$. The source of the
Christoffels $\Gamma$   are two fold: one is
the expansion induced by the motion of the brane inside a warped
background, the other is due to the presence of the black hole.
These modifications to ideal hydrodynamics are not small even at
late stages as far as the strong character of the interaction
sustains. However, these
effects are small far from the fireball center at all times. These
analysis will provide yet another route to non-ideal hydrodynamics
for ${\cal N}$=4 SYM theory at strong coupling. In particular to
the calculation of the transport coefficients. This will be
reported elsewhere.

 \vskip 2cm

\noindent{\large \bf Acknowledgments} \vskip .35cm We thank
Minwalla, Wiseman and especially Aharony for explaining their work
to as well as the intense discussion on the subject following the
appearance of the first version of this paper.
The work of SJS was supported by
KOSEF Grant R01-2004-000-10520-0 and by SRC Program of the KOSEF
with grant number R11 - 2005- 021. The work of ES and IZ was
partially supported by the US-DOE grants DE-FG02-88ER40388 and
DE-FG03-97ER4014.

\vskip 1cm

\end{document}